\documentclass[12pt,a4paper]{article}
\usepackage[top=4cm,bottom=1cm,left=2cm,right=2cm]{geometry}
\usepackage{amsthm}
\usepackage{amsmath}
\usepackage{amssymb}
\usepackage[dvipdfmx,unicode,bookmarks=true, bookmarksnumbered=true, bookmarksopen = false, colorlinks=true]{hyperref}
\usepackage{graphics}
\usepackage{epsfig}
\usepackage{geometry}
\makeatletter \renewenvironment{proof}[1][\proofname]
{\par\pushQED{\qed}\normalfont\topsep6\p@\@plus6\p@\relax\trivlist\item[\hskip\labelsep\bfseries#1\@addpunct{.}]\ignorespaces}{\popQED\endtrivlist\@endpefalse} \makeatother

\theoremstyle{plain}
\newtheorem{thm}{Theorem}
\newtheorem{lem}{Lemma}
\newtheorem{defn}{Definition}
\newtheorem{exmp}{Example}

\newtheorem{prop}{Proposition}

\begin{document}
\author{Yangbo Song\thanks{Department of Economics, UCLA. Email: darcy07@ucla.edu.}
\and Mihaela van der Schaar\thanks{Department of Electrical Engineering, UCLA, and Director of UCLA Center for Engineering Economics, Learning, and Networks. Email: mihaela@ee.ucla.edu.}}
\date{November 20, 2013}
\title{Dynamic Network Formation with Incomplete Information\thanks{We are grateful to William Zame, Ichiro Obara, Moritz Meyer-ter-Vehn, Sanjeev Goyal, Luca Canzian, Simpson Zhang and a number of seminar audiences for suggestions which have significantly improved the paper.}}
\maketitle

\begin{abstract}
How do networks form and what is their ultimate topology?  Most of the literature that addresses these questions assumes $\textit{complete information}$: agents know in advance the value of linking to other agents, even with agents they have never met and with whom they have had no previous interaction (direct or indirect).  This paper addresses the same questions under what seems to us to be the much more natural assumption of $\textit{incomplete information}$: agents do not know in advance -- but must $\textit{learn}$ -- the value of linking to agents they have never met.  We show that the assumption of incomplete information has profound implications for the process of network formation and the topology of networks that ultimately form.  Under complete information, the networks that form and are stable typically have a star, wheel or core-periphery form, with high-value agents in the core. Under incomplete information, the presence of positive externalities (the value of indirect links) implies that a much wider collection of network topologies can emerge and be stable.  Moreover, even when the topologies that emerge are the same, the locations of agents can be very different. For instance, when information is incomplete, it is possible for a hub-and-spokes network with a low-value agent in the center to form and endure permanently: an agent can achieve a central position purely as the result of chance rather than as the result of merit. Perhaps even more strikingly: when information is incomplete, a connected network could form and persist even if, when information were complete, $\textit{no links}$ would ever form, so that the final form would be a totally disconnected network.  All of this can occur even in settings where agents eventually learn $\textit{everything}$ so that information, although initially incomplete, eventually becomes complete.

\begin{flushleft}
\end{flushleft}

\begin{flushleft}
\end{flushleft}
\end{abstract}

\newpage
\section{Introduction}

How do social and economic networks form and what is their ultimate shape (topology)?  The literature, which includes \cite{JW}, \cite{BA}, \cite{BG}, \cite{JWa} and \cite{Ballester}, has addressed these questions in both static and dynamic contexts. The central conclusion of this work is that special shapes of networks can occur and persist. However, this literature makes the strong assumption of \textit{complete information}: agents know \textit{in advance} the value of linking to other agents - even agents they have never met and with whom they have had no previous interaction (direct or indirect).  The present paper addresses the same questions under what seems to us to be the much more natural assumption of \textit{incomplete information}: agents do not know in advance -- but must \textit{learn} -- the values of linking to agents they have never met\footnote{We emphasize that we study learning \textit{during the network formation process}, rather than learning in an exogenously given and fixed network structure. For a study on the latter, see for example \cite{Acemoglu}}. As is usual in environments of incomplete information, agents begin only with \textit{beliefs} about the values of linking to other agents, make choices on the basis of their beliefs, and update their beliefs (learn the true values) on the basis of their experience (history). The network topology is then \textit{endogenously} decided as a result of agents' strategic interaction.

We show that the assumption of incomplete information has profound implications for both the process of network formation and the topology of networks that ultimately form.  When information is complete, the networks that form and persist typically have a star or core-periphery form, with high-value agents in the core.  By contrast, when information is incomplete, a much larger variety of networks and network shapes can form and persist.  Indeed, the set of networks that can form and persist when information is incomplete is a superset (typically a strict superset) of the set of networks that can form and persist when information is complete.  Moreover, even when the network shapes that form are the same or similar, the locations of agents within the network can be very different.  For instance, when information is incomplete, it is possible for a star network with a low-value agent in the center to form and persist indefinitely; thus, an agent can achieve a central position purely as the result of chance rather than as the result of merit.  Perhaps even more strikingly, when information is incomplete, a connected network can form and persist even if, when information were complete, \textit{no links} would ever form so that the final form would be a totally disconnected network.

However, the most important consequence of incomplete information is not that a larger variety of network shapes (topologies) might emerge, but that the particular shape that \textit{does} emerge depends on the history of link formation and of link formation opportunities.  For instance, when information is incomplete, agents $i,j$ might choose to form a link because each \textit{expects} the value of the link to exceed the cost of forming it.  Having formed the link, the agents may learn that their expectations were wrong and so might wish to sever it.  However, before the agents have the opportunity to sever the link, each of them may have formed \textit{other} links, so that the \textit{indirect} value of the link between $i$ and $j$ - the value of the connection to \textit{other} agents - may be sufficiently large that they prefer to maintain the link between them after all.\footnote{Indirect linking provides a positive externality when information is complete as well, but the effect is completely different: $i,j$ will never wish to form a link at some point in time and sever it later.}   However, whether these other links have formed will depend not only on the values of those links but also on the random opportunities presented to form them or not.

We stress that all of this can occur even in settings where agents eventually learn \textit{everything} so that information, although initially incomplete, eventually becomes complete. Incompleteness of information may eventually disappear but its influence may persist forever.

\section{Literature Review}

The literature on network formation studies the characteristics of networks emerging as a result of the strategic interaction of self-interested agents. This literature includes renowned early papers in this area, including \cite{JW}, \cite{BG}, \cite{Watts}, etc. These papers and subsequent works building on them (for instance \cite{JG}) assume that agents are homogeneous, i.e. the value that each agent in the group provides to and receives from another agent is the same. This is the strongest form of \textit{complete information}, in the sense that agents do not only know their exact payoffs from linking to others, but are also aware that the payoffs are solely determined by the network topology, and that the agents' identities play no role in affecting payoff characteristics. As a result, a prominent feature in network topologies that emerge and/or become stable is that they are either \textit{empty} or \textit{connected}. By contrast, in our paper, agents are of different types and thus connecting to them results in \textit{different (heterogeneous)} payoffs; moreover, there is \textit{incomplete information}: an agent does \textit{not} know the types of agents that he has never connected with, but he is able to form \textit{beliefs} based on which he chooses the optimal action. Networks which result under this, more realistic, assumption are strikingly different from those obtained in the model assuming homogeneous agents and complete information. On one hand, connectedness is no longer a key property of stable networks - the formation process converges to connected networks in some range of parameters and to multiple components in other ranges. On the other hand, even if the network stays empty forever under complete information, a non-empty, even connected network may emerge and be stable with positive probability under incomplete information.

In the literature studying how agent heterogeneity affects network formation \cite{HS}, \cite{Galeotti1}, \cite{Galeotti2}, \cite{GG}, \cite{ZV}, \cite{ZV2} \cite{Koenig}, \textit{complete information} is still a common assumption, and the predictions in the above papers are often restricted to a few, specific, types of network topologies such as stars, wheels or core-periphery networks with high-value or low-cost agents enjoying higher connectivity than others. We differ from these works in three aspects. First, as mentioned above, agents have \textit{no precise knowledge} about their exact payoffs due to \textit{incomplete information}; instead, they choose an optimal action according to their \textit{beliefs} about the payoffs that they will obtain from connecting to others. Secondly, we show that the interaction of incomplete information and agent heterogeneity produces a \textit{much wider} range of network topologies, which includes stars, wheels, core-periphery networks, etc. Finally, the topology that emerges and becomes stable strongly depends on the \textit{formation history}: an agent may exhibit a high degree of connectivity in equilibrium not necessarily because he is of a special (high) type but also because initially he was fortunate to obtain sufficiently many links by chance, which in turn attracted others to form and maintain links with him due to the large indirect benefits that he can offer. Therefore, unlike most existing literature that only emphasizes \textit{what} topologies can be formed, we argue that \textit{how} a certain topology comes into being is equally important.

As an important sidenote, we would like to emphasize the difference between this paper and works such as \cite{Koenig}, in which agents choose an optimal level of effort or contribution but the network formation process remains \textit{exogenous}, governed by some commonly known stochastic process. In our model, the network formation process is \textit{endogenous}, in the sense that agents choose directly whether to form links with others. The network topology that emerges then is a result of interaction among individual strategic behavior.

Among the empirical literature on network formation games, works such as \cite{FK}, \cite{CD}, \cite{Goeree} and \cite{RH} have conducted experimental studies on the types of emerging topologies. The experimental results indicate that (1) typical equilibrium network topologies predicted by the existing theoretical analysis are \textit{not} always consistent with the empirical observations; especially, stars are formed only in \textit{a proportion} of the total number of experiments conducted \cite{CD}, \cite{RH}, and such proportions, under some treatments such as a two-way flow of payoffs, are rather low \cite{FK}; (2) even in the experiments where equilibrium network topologies do emerge with high frequency, such topologies are \textit{developed} rather than \textit{born} \cite{Goeree}, which we believe suggests a dynamic network formation process of a sufficiently long duration as a more appropriate environment for stable networks to emerge, compared with a static one. Moreover, the study of networks which actually get formed in large social communities, as presented for instance by \cite{Mele} and \cite{Leung}, shows that in environments where agents are heterogeneous and withhold certain private information in their payoffs from links, numerous phenomena which are not predicted by the existing theoretical literature (such as multiple components and clustering of agents with different attributes) can happen. We believe that incorporating incomplete information in the dynamic network formation game represents a first and important step towards understanding why several previously seemingly irregular network topologies can emerge and remain stable in practice.

The rest of the paper is organized as follows. Section 3 introduces the model. Section 4 analyzes the model in detail and interprets the results. Section 5 discusses an alternative approach in modeling. Section 6 concludes and introduces relevant future research topics.

\section{Model}

Our notation mainly follows \cite{JW}, \cite{BG} and \cite{Watts}.

\subsection{Networks with Incomplete Information}

\subsubsection{Networks and the Agents' Types}

Let $I=\{1,2,...,N\}$ denote a group of $N$ agents. Each agent has a private type $k_i\in X$, where $X$ is a type set. We do not impose any specific assumptions on $X$: it may be finite, countably infinite, or uncountably infinite, as long as the expected payoff on this type set is well-defined (details provided in section 3.2). Let $k_i$ denote agent $i$'s type, and let $\kappa=\{k_i\}_{i=1}^N$ denote the type vector of the agents. The realization of this type vector is drawn i.i.d. for each agent, with prior probability distribution distribution function $H(x)$ (and probability density function $h(x)$\footnote{When $X$ is finite, $h(x)$ denotes its probability mass function.}).

A $\textit{network}$ is denoted by ${\bf{g}}\subset\{ij:i,j\in I,i\neq j\}$, and a $\textit{sub-network}$ of ${\bf{g}}$ on $I'\subset I$, denoted ${\bf{g}}_{sub}(I')$, is defined as a subset of ${\bf{g}}$ such that $ij\in{\bf{g}}_{sub}(I')$ if and only if $i,j\in I'$ and $ij\in{\bf{g}}$. $ij$ is called a $\textit{link}$ between agents $i$ and $j$. We assume throughout that links are $undirected$, in the sense that we do not specify whether link $ij$ points from $i$ to $j$ or vice versa. A network ${\bf{g}}$ is \textit{empty} if ${\bf{g}}=\varnothing$.

We say that agents $i$ and $j$ are $\textit{connected}$, denoted $i\overset{{\bf{g}}}{\leftrightarrow}j$, if there exist $j_1,j_2,...,j_n$ for some $n$ such that $ij_1,j_1j_2,...,j_nj\in{\bf{g}}$. Let $d_{ij}$ denote the $\textit{distance}$, or the smallest number of links between $i$ and $j$. If $i$ and $j$ are not connected, define $d_{ij}:=\infty$. An agent $i$ in a network is a $\textit{singleton}$ if $ij\notin{\bf{g}}$ for any $j\neq i$.

Let $N({\bf{g}})=\{i|\exists j\text{ s.t. }ij\in{\bf{g}}\}$. A $\textit{component}$ of network ${\bf{g}}$ is a maximal connected sub-network, i.e. a set $C\subset {\bf{g}}$ such that for all $i\in N(C)$ and $j\in N(C)$, $i\neq j$, we have $i\overset{C}{\leftrightarrow}j$, and for any $i\in N(C)$ and $j\in N({\bf{g}})$, $ij\in {\bf{g}}$ implies that $ij\in C$. Let $C_i$ denote the component that contains link $ij$ for some $j\neq i$. Unless otherwise specified, in the remaining parts of the paper we use the word "component" to refer to any $\textit{non-empty}$ component.

A network ${\bf{g}}$ is said to be \textit{empty} if ${\bf{g}}=\varnothing$, and \textit{connected} if ${\bf{g}}$ has only one component which is itself. ${\bf{g}}$ is \textit{minimal} if for any component $C\subset {\bf{g}}$ and any link $ij\in C$, $C-ij$ is no longer a component. ${\bf{g}}$ is \textit{minimally connected} if it is minimal and connected.

\subsubsection{Payoff Structure}

Following the assumption in literature on networks that explicitly model non-local externalities\footnote{As opposed to papers that assume pure local externalities, i.e. agents only obtain payoffs from their immediate neighbors. See for example \cite{GG}.}, such as \cite{JW}, \cite{Watts} and \cite{Galeotti2}, we assume that once agents $i$ and $j$ form a link, $i$ not only obtains payoffs from his immediate neighbor $j$, but also from the agents that he is indirectly connected to via that particular link. As stated before, the payoff of an agent also depends on the type vector. Specifically, an agent $i$'s payoff from a network ${\bf{g}}$ is given by 
\begin{align*}
u_i(k_{-i},{\bf{g}})=u_i(k_{-i},C_i):=\sum_{j\overset{C_i}{\leftrightarrow}i}\delta^{d_{ij}-1}f(k_j)-\sum_{j:ij\in C_i}c
\end{align*}
where $f:X\rightarrow \mathbb{R}^{++}$ is the payoff function for an agent from a link with another agent. Therefore, $f(k_j)>0$ denotes the payoff to an agent $i$ by linking to an agent $j$, whose value depends on $j$'s type $k_j$. $c>0$ is the cost of maintaining a link, which is assumed to be bilateral and homogeneous across agents. $\delta\in[0,1]$ denotes a common decay factor\footnote{In this paper, unless otherwise specified, we assume that $\delta\in(0,1)$.}, such that the payoff of $i$ from $j$ with a distance of $d_{ij}$ is $\delta^{d_{ij}-1}f(k_j)$. Note that $c$ is assumed to be independent of the agents' types\footnote{We make this assumption to economize on notations. In the case where costs are also heterogeneous, our analysis and results hold with slightly modified conditions.}.

Let $\mathbb{E}[f(x)]=\int_X f(x)dH(x)$ denote the expected benefit from a link to a single agent, under the prior type distribution. As mentioned before, the only assumption we require on $X$ (and functions $H$ and $f$) is that this expected payoff is well-defined.

\subsection{Dynamic Network Formation Game}

\subsubsection{The Game}

We model the dynamic game in a similar fashion to \cite{Watts} and \cite{JWa}. Time is discrete and the horizon is infinite: $t=0,1,2,...$. The game is played as follows: agents start with an empty network $\varnothing$ in period $0$. In each following period, a pair of agents $(i,j)$ is randomly selected to update the link between them. For simplicity, we assume that the matching probability is uniform, so $(i,j)$ is selected with probability $\frac{2}{N(N-1)}$. We assume that when agents $i$ and $j$ are selected, $i$ can observe $C_j$, and vice versa. The two agents then play a simultaneous move game, where each can choose to sever the link between them if there is one, and if there is not, whether to agree to form a link with the other agent. Let $a_{ij}=1$ denote the action that $i$ agrees to form a link with $j$ (if there is no existing link) or not to sever the link (if there is an existing one), and $a_{ij}=0$ otherwise. A link is formed or maintained after bilateral consent (i.e. $a_{ij}=a_{ji}=1$). Let $\gamma(t):=\{(i,j)_{t'}\}_{t'=1}^{t}$ ($t\geq 1$) denote a $\textit{selection path}$ up to time $t$, or the set of selected pairs of agents, ordered from $1$ to $t$. The agents are assumed to be myopic, i.e. they only care about their current period payoffs.

\subsubsection{Updating Rule}

With incomplete information, agents maximize their $\textit{expected payoffs}$, rather than actual payoffs, when deciding the optimal action. Therefore, the \textit{belief} of an agent on the types of the other agents plays a crucial role in shaping his behavioral patterns. For any $i$, we let $B_i\in\Delta X^{N}$ denote agent $i$'s belief on the type vector of the group, and $B=(B_1,...,B_N)$ define a belief vector for the group of agents. Note that for any $i$, $B_i$ must put zero probability on any type vector where $i$'s type differs from the true type. Moreover, we assume the following simple updating rule for the agents:
\begin{itemize}
\item{1.} If two agents are ever connected, they know each other's type.
\item{2.} Otherwise, their belief on each other's type remains at the prior.
\end{itemize}
In other words, agents can only observe their own formation history: they do not observe the actions by agents in other components and thus make no relevant inferences on their types. A similar assumption appears in \cite{McBride}, which is denoted as imperfect monitoring and describes agents' inability to observe all other agents' strategies in a static network formation game. A plausible alternative updating rule is to allow agents to observe each other's past actions and perform Bayesian updating accordingly, which results in very complicated belief formation. We will discuss this in a later section. Throughout the paper, we assume that the updating rule is common knowledge among the agents.

\section{Analysis}

In this section, we analyze the nature of the network formation process and show a clear contrast between results under complete information and incomplete information. We begin by defining the solution concept for the two-player game each period, and the notion of a stable network.

\subsection{Stable Optimistic Equilibrium and Stable Network}

\subsubsection{Stable Optimistic Equilibrium}

Let $\mathcal{C}$ denote the set of all possible components (including the empty component), i.e. when agents $i$ and $j$ are selected, for agent $i$, $\mathcal{C}$ is the set of all possible observations of $C_j$. Let $\mathcal{B}$ denote the set of all possible beliefs for an agent. Denote $\mathcal{K}=X\times \mathcal{C}\times \mathcal{B}$ as the $\textit{set of information}$\footnote{Technically speaking, agent $i$'s information also includes $C_i$, but omitting such information does not affect the existence and any property of the equilibrium we define below.} for an agent who is selected to update a link. The interpretation of knowledge is the following: the first argument is the agent's own type, the second argument is the component containing his counterparty, i.e. the other agent selected, and the third argument is his current belief on the type vector.

Now we define an agent's (pure) strategy in the two-player game after a pair of agents is selected, followed by our proposed solution concept, which we call a \textit{stable optimistic equilibrium (SOE)}:

\begin{defn}
Agent $i$'s (pure) strategy towards $j$ is $s_{ij}:\mathcal{K}\rightarrow\{0,1\}$, a function from agent $i$'s set of information to his set of actions.
\end{defn}

\begin{defn}
A strategy profile $s=(s_{ij},s_{ji})$ is a stable optimistic equilibrium (SOE) if: (1) it is mutual best response; (2) $s_{ij}=1$ if
\begin{align*}
\mathbb{E}[u_i(k_i,k_{-i},(C_i\cup C_j)+ij)|B_i]\geq u_i(k_i,k_{-i},C_i)
\end{align*}
\end{defn}

In words, a SOE is an equilibrium where any agent would choose to agree to form a link (if there is no existing link) or choose not to sever the link (if there is an existing one) as long as the expected payoff from the link is non-negative. It is $\textit{stable}$ because the prescribed strategy is robust to small probabilistic changes in the counterparty's strategy when the above inequality is strict. More specifically, if some agent $j$ other than $i$ changes $s_{ji}$ with a sufficiently small probability $\epsilon$, $i$'s best response would not change. It is \textit{optimistic} in the sense that it excludes the ``pessimistic'', or null equilibria in which no link formation occurs even though each agent has a non-negative expected payoff from the potential link\footnote{In fact, it is easy to see that except in the cases with indifference, there cannot be any equilibrium in which a link is formed with positive probability and agents use type-dependent strategies. Therefore, in most cases the SOE and the null equilibria are the only two categories of equilibria.}. In other words, agents choose to link when they are at least indifferent. The following lemma shows the existence and uniqueness of such an equilibrium.

\begin{lem}
In every period, for any pair of selected agents, any network topology and any belief vector according to the simple updating rule, a SOE exists and is unique.
\end{lem}

\begin{proof}
See Appendix.
\end{proof}

Lemma 1 together with the above robustness property of a SOE ensures that the outcome of the formation process is unique and robust to small perturbation.
\\

\subsubsection{Stable Network}

Let ${\bf{g}}(\gamma(t))$ denote the unique network formed after period $t$, following a selection path of $\gamma(t)$. Let $B(\gamma(t))$ denote the associated belief vector after period $t$. By Lemma 1 we know that both ${\bf{g}}(\gamma(t))$ and $B(\gamma(t))$ are well-defined. Let $\sigma(t):=(\gamma(t),\{{\bf{g}}(\gamma(t'))\}_{t'=1}^{t})$ denote a $\textit{formation history}$ up to time $t$. We say that:
\begin{itemize}
\item{1.} A network together with the associated beliefs $({\bf{g}},B)$ $\textit{can emerge}$ if there exists a selection path $\gamma(t)$ for some $t$, such that ${\bf{g}}={\bf{g}}(\gamma(t))$, $B=B(\gamma(t))$.
\item{2.} $({\bf{g}},B)$ is a $\textit{stable network}$ if no link is formed or broken given any subsequent selection path.
\item{3.} The formation process $\textit{can converge to}$ ${\bf{g}}$ if there exists a selection path $\gamma(t)$ for some $t$, such that ${\bf{g}}={\bf{g}}(\gamma(t))$ and $({\bf{g}}(\gamma(t)),B(\gamma(t)))$ is stable. Mathematically, we denote it as $\lim_{t\rightarrow\infty}{\bf{g}}(\gamma(t))={\bf{g}}$ for some fixed $\gamma(\infty)$.
\end{itemize}

\subsection{Contrast with Complete Information}

Before discussing the differences between the network topologies that can emerge and be stable under complete information and incomplete information, we first inspect how long incomplete information can persist. We say that \textit{information is complete} when every agent's belief is the degenerate belief on the true type vector $\kappa$, i.e. $Prob(\kappa|B_i)=1$ for any $i$, and that \textit{information is incomplete} otherwise.

\begin{prop}
For any $\kappa$:
\begin{itemize}
\item{1.} If $\mathbb{E}[f(x)]< c$, information is never complete: agents' beliefs stay at the prior.
\item{2.} If $\mathbb{E}[f(x)]\geq c$, information becomes complete within finitely many periods almost surely, and information is complete in any stable network.
\end{itemize}
\end{prop}

\begin{proof}
See Appendix.
\end{proof}

In essence, when the expected benefit from a link under the prior exceeds the link formation cost, everyone eventually learns the true type vector with probability $1$ over time. Indeed, as agents are willing to form links with others of unknown type, after sufficiently many periods the probability of unconnected agents still existing in the group would be arbitrarily small. In other words, the effect of incomplete information and thus the difference with complete information only occurs in an early stage of the formation process.

Nevertheless, even though Proposition 1 may leave the impression that incomplete information is not so important as it only takes effect in the short run, we will emphasize in the following analysis that such short-term influence is actually persistent over time.

Let $G_C(\kappa)=\{{\bf{g}}:{\bf{g}}={\bf{g}}(\gamma(t))\text{ for some $\gamma(t)$}\}$ denote the set of networks that can emerge under complete information given $\kappa$, and $G_{IC}(\kappa)$ that under incomplete information. Similarly, let $G_C^*(\kappa)=\{{\bf{g}}\in G_C(\kappa):{\bf{g}}=\lim_{t\rightarrow\infty}{\bf{g}}(\gamma(t))$ for some $\gamma(\infty)\}$ denote the set of networks that can emerge and be stable under complete information given $\kappa$, and $G_{IC}^*(\kappa)$ that under incomplete information.

\begin{thm}
For any $\kappa$:
\begin{itemize}
\item{1.} If $\mathbb{E}[f(x)]< c$, $G_{IC}(\kappa)=G_{IC}^*(\kappa)=\{\varnothing\}$.
\item{2.} If $\mathbb{E}[f(x)]\geq c$, $G_{IC}(\kappa)\supset G_C(\kappa)$, and $G^*_{IC}(\kappa)\supset G^*_C(\kappa)$.
\end{itemize}
\end{thm}

\begin{proof}
If $\mathbb{E}[f(x)]< c$: as in the proof of Proposition 1, no link would ever form and thus the only network that can emerge (and be stable) is the empty network.

If $\mathbb{E}[f(x)]\geq c$: for any ${\bf{g}}\in G_C(\kappa)$:

If ${\bf{g}}$ is empty: since ${\bf{g}}\in G_C(\kappa)$, there must exist two agents $i,j$ such that $f(k_i)<c$ or $f(k_j)<c$. Consider the selection path $\gamma(2)=((i,j)_1,(i,j)_2)$ under incomplete information. It is clear that a link would form between $i$ and $j$ in period $1$, but the link would then be severed in period $2$, and thus ${\bf{g}}={\bf{g}}(\gamma(2))$, which implies that ${\bf{g}}\in G_{IC}(\kappa)$.

If ${\bf{g}}$ is non-empty: consider any selection path $\gamma_C(t)$ such that ${\bf{g}}$ emerges for the first time in period $t$. By the
definition of $G_C(\kappa)$, we know that such $\gamma_C(t)$ exists. Let $\gamma_{IC}(t')$ be a selection path constructed from $\gamma_C(t)$ such that the pairs of agents in $\gamma_C(t)$ between whom there is no existing link, but a new link is not formed either, are deleted.

Consider the formation process under incomplete information given $\gamma_{IC}(t')$. By the simple updating rule, we know that given the same network structure, if a link is formed between $i$ and $j$ under complete information, it will also be formed under incomplete information regardless of whether $i$ and $j$ know each other's type. Also, it is clear that the decision of severing a link by any agent is the same under complete information and incomplete information, since such a decision is based on the realized payoff. Therefore, the formation process yields the same link formation and severance results under complete information given $\gamma_C(t)$ and under incomplete information given $\gamma_{IC}(t')$. Hence given $\gamma_{IC}(t')$, ${\bf{g}}$ emerges for the first time in period $t'$ under incomplete information, which implies that ${\bf{g}}\in G_{IC}(\kappa)$. Therefore $G_C(\kappa)\in G_{IC}(\kappa)$.

By the above argument, we already know that any network that can emerge under complete information can also emerge under incomplete information. Thus it suffices to show that, for any network ${\bf{g}}\in G_C^*(\kappa)$, there exists a subsequent selection path under incomplete information after ${\bf{g}}$'s first appearance that would make ${\bf{g}}$ stable. We prove the result by construction. 

If ${\bf{g}}$ is empty: consider the selection path $\gamma(\frac{N(N-1)}{2})$ such that every pair of agents is selected exactly twice consecutively. Since by assumption ${\bf{g}}\in G_C^*(\kappa)$, we know that for every pair of agents a link would first form and then be severed in the next period. In period $\frac{N(N-1)}{2}+1$, information is complete and the empty network becomes stable.

If ${\bf{g}}$ is non-empty: denoting the number of components in ${\bf{g}}$ as $q({\bf{g}})$, under incomplete information let the subsequent selection path after ${\bf{g}}$'s first appearance be such that in the first $q({\bf{g}})(q({\bf{g}})-1)$ periods, two agents from different components are selected exactly twice consecutively and every two components are involved. Since by assumption ${\bf{g}}\in G_C^*(\kappa)$, which means that ${\bf{g}}$ is stable under complete information, when a pair of agents from different components is selected for the second time, either there is no existing link between them and no link would be formed, or an existing link would be severed. In either case, the agents know each other's type as well as the types of agents in the counterparty's component. Therefore, after $q({\bf{g}})(q({\bf{g}})-1)$ periods, every agent knows $\kappa$ and essentially there is no incomplete information. Again by the assumption that ${\bf{g}}\in G_C^*(\kappa)$, we can conclude that g is such that no link would be formed or severed in any later period given any selection path. Thus ${\bf{g}}\in G_{IC}^*(\kappa)$, which completes the proof.
\end{proof}

Theorem 1 states that when expected benefits are sufficiently high, if some network can emerge (and be stable) under complete information, then it can also do so under incomplete information. Intuitively, if a link could be formed under complete information, then given high expected benefits and the simple updating rule, it can also be formed under incomplete information, whether or not the relevant agents know each other's type. Note that the reverse is not necessarily true: even if a link could be formed under incomplete information, it may not form under complete information because the expected payoffs are sufficiently higher than the realized payoffs, i.e. in the incomplete information setting, were the agents to know each other's type beforehand, the link may never be formed. In other words, under incomplete information, high expected payoffs can initialize link formation such that even though agents would ``regret'' the links they form after knowing each other's type, more links have formed before they are selected again to update their initial links. Then due to increasing returns to link formation, the positive externalities would in turn ensure that the initial links are maintained. The following example illustrates this point.

\begin{exmp}
Assume the following: $N=5$, $X=\{a,b\}$, $k_i=b$ $\forall i=1,...,5$, and the other parameters are such that $f(b)<c$, $\mathbb{E}[f(x)]\geq c$, $(1+\delta-\delta^2-\delta^3)f(b)\geq c$.

Consider the following selection path: $(1,2)$, $(2,3)$, $(3,4)$, $(4,5)$, $(1,5)$. Under complete information, it is clear that the network remains empty regardless of the selection path, as in Figure 1(A). Under incomplete information, the SOE can be explicitly computed in each period. For example, in period 1, agent 1's expected payoff from the link with agent 2 is $\mathbb{E}[f(x)]\geq c$ and vice versa, and thus the link is formed. The formation process is shown in Figure 1(B).

\begin{figure}[h]
\centering
\includegraphics[width=5.5in]{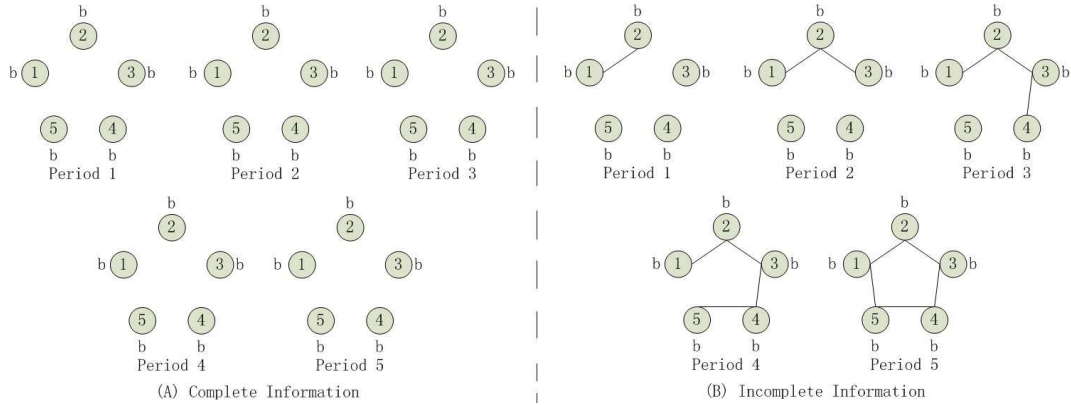}
\caption{Empty under Complete Info. vs. Connected under Incomplete Info.}
\end{figure}

According to the assumptions on parameters, one can then easily show that the network formed in period 5 is stable.
\end{exmp}

Another feature of incomplete information is the \textit{history dependence} of the formation process, in the sense that the ultimate network topology depends greatly on the selection path. As a result, even if a type is more valuable or preferable than the other, under incomplete information it is not necessary that an agent of that type ends up with a higher connectivity degree. Consider the following example: assume the same parameter values as in Example 1, and consider a group of agents consisting of $4$ type $a$ agents and $5$ type $b$ agents. There exists a selection path such that: under complete information, the formation process converges to a star network with only type $a$ agents (Figure 2(A)); under incomplete information, the formation process converges to a "hub-and-spokes" network (Figure 2(B))\footnote{One such selection path is $(1,2)$, $(1,3)$, $(2,6)$, $(3,7)$, $(6,,7)$, $(1,5)$, $(5,9)$, $(6,9)$, $(1,4)$, $(4,8)$, $(7,8)$, $(8,9)$, $(2.5)$, $(3,5)$, $(4,5)$.}.

\begin{figure}[h]
\centering
\includegraphics[width=5in]{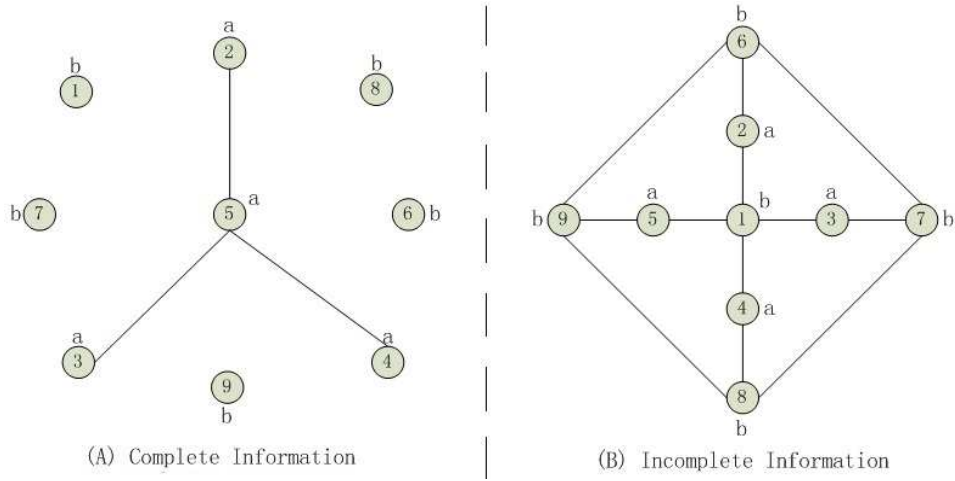}
\caption{Different Connectivity Degree Distributions}
\end{figure}

Under complete information, the center of the star network has to be a type $a$ agent since no type $b$ agent ever gets linked with anyone else. By contrast, under incomplete information, it first becomes possible for two type $b$ agents to form a link; and then, as it turns out in this particular topology, each type $b$ agent's distance with the type $a$ agent is sufficiently small. Even though the type $a$ agent has a low connectivity degree, the other agents do not find a new link with the type $a$ agent attractive, because it does not offer sufficient indirect benefits. Hence, the one agent with the more valuable type -- type $a$ -- ends up with the lowest connectivity degree in the network. This is in stark contrast with the existing results in the literature (for instance \cite{GG}), which often show that a more valuable agent is better connected. This example also highlights and clarifies the point made by Theorem 1: incomplete information generates a superset of \textit{networks}, not \textit{links}, as compared with complete information. In other words, \textit{new and different} networks can be formed under incomplete information, rather than a mere addition of links to networks formed under complete information. Indeed, the network in Figure 2(A) has 3 links and that in Figure 2(B) has 12, but they share \textit{no links in common}.

Also, even when $\mathbb{E}[f(x)]\geq c$, incomplete information does \textit{not} imply a greater number of links. For instance, let $X=\{a,b\}$, and consider a group of 8 type $a$ agents (indexed 1, 2, $\cdots$, 8) and 1 type $b$ agent (indexed 9). The payoffs are $f(b)<c$, $f(a)\geq c$, $(1-\delta)f(a)<c\leq (1-\delta^2)f(a)$ and $\mathbb{E}[f(x)]\geq c$. Let the selection path be as follows: first, select 9 once with each of 1-8. Then select 12, 23, $\cdots$, 78, 81. Finally, select 15, 26, 37, 48. The convergence result is shown in Figure 3 below: under complete information, the network has 12 links, while under incomplete information it has only 8.

\begin{figure}[h]
\centering
\includegraphics[width=5in]{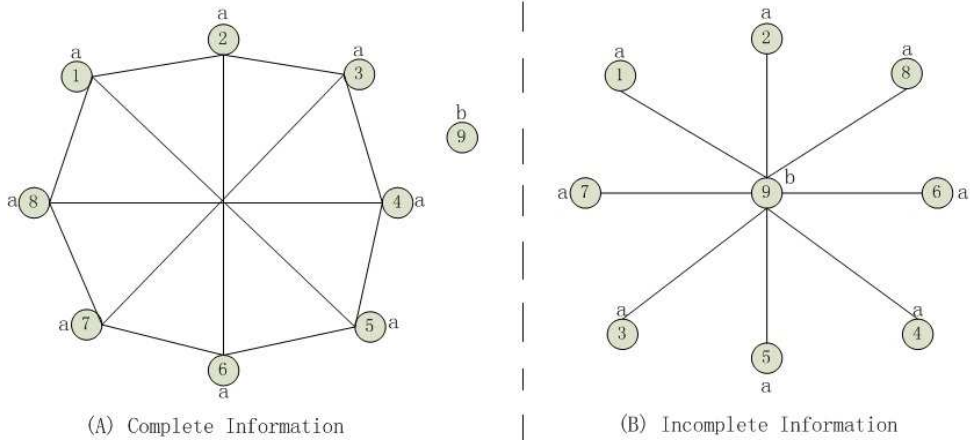}
\caption{More Links under Complete Information}
\end{figure} 

Furthermore, under incomplete information, the probability that the formation process converges to a different network topology from those under complete information can be significant. The following figure shows simulation results\footnote{We assume for the simulation that $\delta=0.8$ and $p=0.5$. For each value of $N$, we run $500$ simulations and average the results. Each simulation consists of a fixed long run of $5N(N-1)$ periods, so that each link is selected for $10$ times on average.}. Under complete information, no type $b$ agent would ever get linked; under incomplete information, the frequency that some type $b$ agent remains linked by the end rises to more than $0.8$ for a range of the group size $N$.

\begin{figure}[h]
\centering
\includegraphics[width=5in]{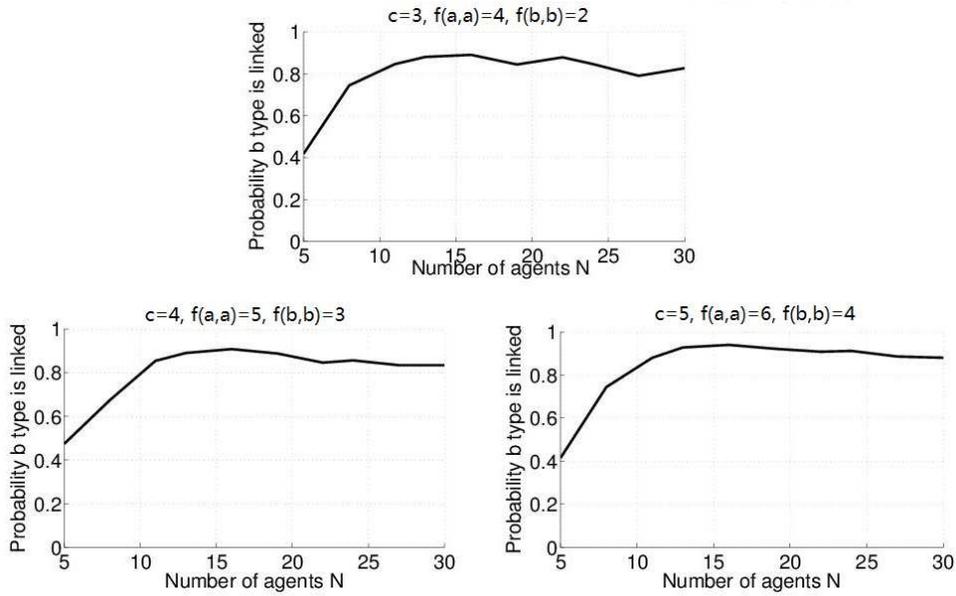}
\caption{Simulation: Significant Probability of Difference}
\end{figure}

The above examples also indicate that, unlike in most existing literature that specifies only a few types of network topologies as the only possible ones in equilibrium, under incomplete information the formation process can converge to potentially $\textit{more}$ $\textit{types}$ of networks, as there may exist some network ${\bf{g}}$ which can emerge and be stable under incomplete information, but not under complete information. We define some typical network structures in the literature and the empirical works below:

\begin{itemize}
\item{1.} ${\bf{g}}$ is $\textit{complete}$ if $ij\in{\bf{g}}$ $\forall i,j$ such that $i\neq j$. 
\item{2.} ${\bf{g}}$ is a $\textit{star}$ network if there exists $i\in I$ such that $ij\in{\bf{g}}$ $\forall j\neq i,j\in I$ and $i'j\notin {\bf{g}}$ $\forall i',j\neq i$.
\item{3.} ${\bf{g}}$ is a $\textit{core-periphery}$ network if there exists non-empty $I'\subsetneq I$, such that $ij\in{\bf{g}}$ $\forall i,j\in I',i\neq j$, and that $\forall j'\in I\setminus I'$, $ij'\in{\bf{g}}$ for some $i\in I'$ and $jj'\notin{\bf{g}}$ $\forall j\neq i$. Note that a star network is a special case of a core-periphery network.
\item{4.} ${\bf{g}}$ is a $\textit{tree}$ network if there exists a partition of $I$, $I_1,...,I_n$, such that (1)$\#(I_1)=1$; (2)$\forall n'=2,...,n$, each agent in $I_{n'}$ has one and only one link with some agent in $I_{n'-1}$; (3)no other link exists.\footnote{Essentially, a tree network is equivalent to a minimally connected network, and a star network is a special case of a tree network.}
\item{5.} ${\bf{g}}$ is a $\textit{wheel}$ network if there exists a bijection $\pi:I\rightarrow I$ such that ${\bf{g}}=\{\pi^{-1}(1)\pi^{-1}(2),$ $\pi^{-1}(2)\pi^{-1}(3),$ $...,$ $\pi^{-1}(N-1)\pi^{-1}(N)\}$.
\end{itemize}

\begin{thm}
Assume that $\mathbb{E}[f(x)]\geq c$. For any $\kappa$, if a network ${\bf{g}}$ is stable and belongs to one of the following categories:
\begin{itemize}
\item{1.} Empty network;
\item{2.} Minimally connected network (i.e. tree network, including star network);
\item{3.} Fully connected network;
\item{4.} Core-periphery network;
\item{5.} Wheel network.
\end{itemize}
then ${\bf{g}}\in G_{IC}^*(\kappa)$.
\end{thm}

\begin{proof}
By the assumption that ${\bf{g}}$ is connected and stable, it suffices to show that when ${\bf{g}}$ belongs to any of the categories there exists a selection path such that ${\bf{g}}$ can emerge in the formation process. We discuss case by case and prove by construction below.

1: See the proof of Theorem 1.

2: let $L$ be the total number of links in ${\bf{g}}$. Let the selection path be such that the pair of agents for each link in ${\bf{g}}$ is selected once and only once in the first $L$ periods. Since $\mathbb{E}[f(x)]\geq c$, we know that each link will be formed, and thus ${\bf{g}}$ emerges in period $L$.

3: Since ${\bf{g}}$ is fully connected and stable, we know that for any two agents $i$ and $j$, $(1-\delta)f(k_i)\geq c$. Therefore regardless of the selection path ${\bf{g}}$ would emerge.

4: Let the selection path be such that: first each periphery agent is selected once and only once with their corresponding core agent, then every two core agents are selected once and only once before any other pair of agents is selected. By Assumption 2 and the assumption that ${\bf{g}}$ is stable, we know that each link will be formed, and thus ${\bf{g}}$ emerges after the last pair of core agents is selected.

5: Let the selection path be such that the pair of agents for each link in ${\bf{g}}$ is selected once and only once in the first $N-1$ periods. By Assumption 2 and the assumption that ${\bf{g}}$ is stable, we know that each link will be formed, and thus ${\bf{g}}$ emerges in period $N-1$.
\end{proof}

Theorem 2 explicitly characterizes types of connected networks that can emerge and be stable under incomplete information, and most typical networks in both the literature and empirical studies are included. However, note that there may be some stable networks that cannot emerge under complete or incomplete information -- e.g. networks with links such that (1) the benefit from any one link cannot cover the maintenance cost without the existence of the other links and (2) the network would still be connected if these links were severed. Such network topologies may never be formed since only one pair of agents is selected in each period, and the agents are myopic.

\subsection{Characterizing Topological Differences}

In the previous analysis, we have seen that even with the same selection path, very different networks can emerge and be stable under incomplete information; in this section, we formalize a way of describing such topological differences, and characterize the corresponding conditions under which these differences are achieved.

Consider a selection path such that the formation process converges under both complete and incomplete information. The following lemma establishes its existence.

\begin{lem}
For any $\kappa$, there always exists a $\gamma(\infty)$ such that the formation process converges under both complete and incomplete information.
\end{lem}

\begin{proof}
See Appendix.
\end{proof}

We say that the difference between the network topologies is \textit{minimal} if the formation process converges to the same network, and \textit{maximal} if at least one of the networks is non-empty, and they have no common links. In addition, we say that $i$ is a \textbf{low-value} agent if $f(k_i)<c$, a \textbf{medium-value} agent if $(1-\delta)f(k_i)<c\leq f(k_i)$ and a \textbf{high-value} agent if $(1-\delta)f(k_i)\geq c$. Let $n_l$, $n_m$ and $n_h$ denote the number of agents in the corresponding category.

\begin{prop}
For any $\kappa$, the following properties hold:
\begin{itemize}
\item{1.} If $\mathbb{E}[f(x)]<c$, then the minimal difference can be achieved if $n_m+n_h\leq 1$, and the maximal difference can be achieved otherwise.
\item{2.} If $\mathbb{E}[f(x)]\geq c$, then the minimal difference can always be achieved, and:
\begin{itemize}
\item{a.} If $n_h\geq 2$ or $n_l=0$, then the maximal difference cannot be achieved.
\item{b.} If $n_h<2$ and $n_l>0$, then the maximal difference can be achieved if (1) $n_m+n_h$ is sufficiently large, or (2) $n_m+n_h\geq 2$, $\delta$ is sufficiently close to 1 and $n_l$ is sufficiently large.
\end{itemize}
\end{itemize}
\end{prop}

\begin{proof}
See Appendix.
\end{proof}

Case 2(b) in the above proposition is of particular interest, because apart from achieving the maximal difference, it also highlights the particular types of network topology that can result in such a difference. When $n_m+n_h$ is sufficiently large, a \textit{star} network with a low-value agent can emerge and be stable under incomplete information, which immediately implies that there are no common links with any stable network under complete information. When $n_l$ is sufficiently large, a \textit{line} network (i.e. a tree network with only one agent in each subset in the partition of $I$) under incomplete information, where low-value agents and medium-value or high-value agents are linked alternately, will ensure the maximal difference.

\subsection{Social Welfare}

An alternative and very important way of comparing complete and incomplete information is to evaluate the upper bound in social welfare in the two cases. Formally, let $W_C(\kappa)$ and $W_{IC}(\kappa)$ be the maximum social welfare that can be achieved by a network in $G_C^*(\kappa)$ and $G_{IC}^*(\kappa)$ respectively. By Theorem 1, it is clear that under incomplete information, if $\mathbb{E}[f(x)]\geq c$, this welfare upper bound is weakly higher than that under complete information, but we aim further at characterizing conditions under which the maximum social welfare in one case is strictly higher than, equal to, or strictly lower than that in the other case.

\begin{lem}
Under both complete and incomplete information, if some stable network ${\bf{g}}_1$ is a proper superset of some other stable network ${\bf{g}}_2$, then every agent's payoff is weakly higher in ${\bf{g}}_1$ than in ${\bf{g}}_2$. As a result, ${\bf{g}}_1$ yields a weakly higher social welfare than ${\bf{g}}_2$.
\end{lem}

\begin{proof}
See Appendix.
\end{proof}

This lemma clarifies the social welfare relation between two stable networks when one contains the other. Then we can show the following result:

\begin{prop}
For any $\kappa$, the following properties hold:
\begin{itemize}
\item{1.} If $\mathbb{E}[f(x)]<c$, then $W_C(\kappa)=W_{IC}(\kappa)$ if $n_m+n_h\leq 1$, and $W_C(\kappa)>W_{IC}(\kappa)$ otherwise.
\item{2.} If $\mathbb{E}[f(x)]\geq c$, then $W_C(\kappa)\leq W_{IC}(\kappa)$, and:
\begin{itemize}
\item{a.} If $n_l=0$, then $W_C(\kappa)=W_{IC}(\kappa)$.
\item{b.} If $n_l>0$ and $n_m+n_h=1$, then $W_C(\kappa)<W_{IC}(\kappa)$ if there exists a stable wheel network among a subset of the agents.
\item{c.} If $n_l>0$ and $n_m+n_h>1$, then $W_C(\kappa)<W_{IC}(\kappa)$ if $\delta$ is sufficiently close to $1$. 
\end{itemize}
\end{itemize}
\end{prop}

\begin{proof}
See Appendix.
\end{proof}

Just as Proposition 2, this result points to particular network topologies (2(b) and 2(c)) that bring about a clear welfare comparison. When $n_l>0$ and $n_m+n_h=1$, the empty network is the only stable one that can emerge under complete information; for any other network that can emerge under incomplete information to be stable, the network must exhibit a "wheel-like" feature, i.e. apart from the medium-value or high-value agent, every agent must have at least two links. And once such a network is stable, it can be immediately shown that it yields a strictly positive social welfare. When $n_l>0$ and $n_m+n_h>1$, as $\delta$ gets sufficiently close to $1$ the network that yields the highest social welfare must be minimal; then under incomplete information, there always exists a way to "insert" a low-value agent between two medium-value or high-value agents, which brings almost no change to the payoffs of the medium-value or high-value agents (since $\delta$ is close to 1) but generates a strictly positive payoff for the low-value agent. Therefore, social welfare is strictly improved.

\section{Alternative Updating Rules}

The results we have derived so far are based on the simple updating rule, which assumes that every agent's posterior belief on another agent's type is binary: either it is the degenerate belief on the true type, or the prior. Note again that such an updating rule implicitly assumes that agents can only observe their own formation history. If the agents adopt a different updating rule, which reflects either more or less available information, the formation process can exhibit a much different pattern. We discuss one such alternative in detail, which we call $\textit{Bayesian learning by formation history}$.

We assume that agents can observe the entire formation history, i.e. the pair of agents selected and the resulting network structure each period, in addition to knowing the types of agents connected to themselves. They then apply Bayesian updating in forming posterior beliefs. The following result highlights the key difference between the simple learning rule and this alternative.

\begin{prop}
Under Bayesian learning by formation history, assume that in the prior type distribution, the probability of an agent being low-value is positive. Then for any agent $i$ of type $k_i$, when there are sufficiently many low-value agents, there exists a formation history $\sigma(t)$ for some $t$ such that $\mathbb{E}[f(k_i)|\sigma(t)]<c$.
\end{prop}

\begin{proof}
See Appendix.
\end{proof}

A major implication here is that Bayesian learning by formation history makes it possible for the posterior probability of an agent being of high type to fall close to $0$. As a result, even if making a link with some agent $i$ is incentivized with the simple learning rule, it may no longer be the case under Bayesian learning by formation history, given some particular selection path that would drag posterior beliefs towards $i$ being of type $b$. The following example illustrates this difference.

\begin{exmp}
Assume the following: $N=5$, $X=\{a,b\}$, $k_i=b$ $\forall i=1,...,5$, and the other parameters are such that $f(b)<c$, $\mathbb{E}[f(x)]\geq c$, $(1+\delta-\delta^2-\delta^3)f(b)\geq c$. Let $p=h(a)$, and assume that $\frac{p(1-p)}{1-p^2}f(a)+\frac{p(1-p)+(1-p)^2}{1-p^2}f(b)<c$. Consider the following selection path in period 1-9: $(1,3)$, $(1,3)$, $(2,4)$, $(2,4)$, $(1,2)$, $(3,4)$, $(4,5)$, $(2,3)$, $(1,5)$. Note that in this environment with pure peer effect, the existence of a SOE (and thus the uniqueness) is still valid under Bayesian learning by formation history, as any agent's expected payoff from any link is not affected by his own type.

Under the simple learning rule, the formation process is shown in Figure 5(A). The network formed in period 9 is stable; under Bayesian learning by formation history, the formation process is shown in Figure 5(B). The network formed in period 4 is stable.
\begin{figure}[h]
\centering
\includegraphics[width=5.5in]{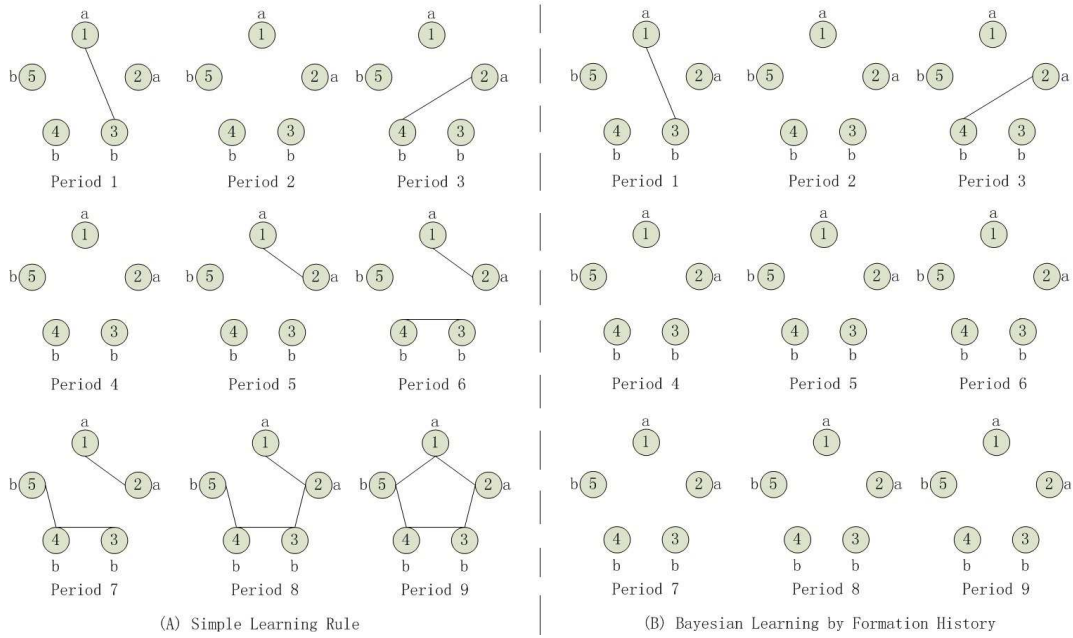}
\caption{Connected under Simple Updating Rule vs. Empty under Bayesian Learning by Formation History}
\end{figure}

Here with the simple learning rule, agents hold the prior belief each time they are selected with another agent with an unknown type, and thus the given selection path induces a connected network at last. Yet with Bayesian learning by formation history, each agent updates from their observation to conclude that others are low-value with a sufficiently large probability, and thus are unwilling to make any link.
\end{exmp}

One implication of the above proposition and example is that more learning can sometimes be ``bad'', i.e. it may lead to inefficient outcomes. Despite the specific differences brought about by an alternative updating rule, our general results still hold under a range of parameters. For instance, it can be shown that if any typical network as depicted in Theorem 2 can emerge and be stable under complete information, then it can under incomplete information and Bayesian learning by formation history as well.

\section{Conclusion and Future Research}

In this paper we analyzed the network formation process under agent heterogeneity and incomplete information. Our results are in stark contrast with the existing literature: instead of restricting the equilibrium network topologies to fall into one or two specific categories, our model generates a great variety of network types. Besides \textit{what} networks can emerge as a result of convergence, we argue that it is also important to understand \textit{how} a network gets formed, since we want to know, for instance, \textit{why} some agents become central and others do not. While link formation and belief formation are usually treated as two independent processes to be studied separately, we combine them in our model and show that belief formation is in fact a key factor that could facilitate or deter link formation. Even if incomplete information vanishes in the long run, its impact on shaping the network topology is persistent.

Several future research topics can be built up on the basis of our model. One of these challenges is to pin down the structure of an efficient network and implement it in a game-theoretic setting. The usual definition on efficiency in networks adopted in the literature is $\textit{strong efficiency}$, i.e. a network is strongly efficient if it maximizes the sum of agents' payoffs. In general, we know that a strongly efficient network must exist (though not necessarily be unique) because the set of possible network structures is finite. However, since payoffs are heterogeneous across agents according to the type vector, the exact topology of an efficient network becomes difficult to characterize; moreover, the efficient network may not be unique because in an agent-heterogeneous environment there could be multiple ways of generating the same level of social welfare.

Most importantly, we have assumed throughout, as does most of the literature, that agents are myopic rather than forward-looking. If it is assumed otherwise that agents are foresighted and are concerned about both their current and future welfare, then the aim of analysis essentially becomes solving an agent's dynamic optimization problem in the presence of other similarly foresighted agents. One can then easily anticipate a very different evolution pattern of network topologies as well as very different stable network topologies in the limit, for now link formation does not only serve as an action of maximizing the current expected payoff, but also as a way of acquiring information for potential future benefit. It is our conjecture that link formation would be more frequent at least in an early stage of the formation process, since on top of the incentives already discussed in this paper, agents may also be willing to endure current payoff losses in return for valuable information about the type vector.

\appendix
\section*{APPENDIX}

\begin{proof}[Proof of Lemma 1]
Assume that agents $i$ and $j$ are selected in some period. Based on the simple updating rule, either $i$ and $j$ know each other's type, or their belief on each other's type remains at the prior. If they know each other's type, then since actions are binary for each agent, the strategy described in a SOE exists and is unique for each agent. It is also clear that the strategy is a best response for each agent since it is a weakly dominant strategy.

If the agents do not know each other's type, and hence their belief remains at the prior, then if $\mathbb{E}[f(x)]\geq 0$, it is indeed mutual best response for the agents to agree regardless of their own types; similarly, if $\mathbb{E}[f(x)]< 0$, it is again mutual best response for the agents not to agree regardless of their own types. Therefore, the strategy profile described in a SOE is indeed an equilibrium. Finally, it is clear that such a profile is unique.
\end{proof}

\begin{proof}[Proof of Proposition 1]
If $\mathbb{E}[X]<c$: by inspecting the SOE we know that for any pair of agents selected, no link would be formed in any period. Therefore, no agent ever learns the type of any other agent, and the beliefs would stay at the prior.

If $\mathbb{E}[X]\geq c$: we first show that information becomes complete within finitely many periods almost surely. It suffices to show that any two agents are connected for at least one period within finitely many periods almost surely. Since $\mathbb{E}[X]\geq c$, by the definition of SOE it further suffices to show that any two agents are selected at least once within finitely many periods almost surely. Consider any two agents $i$ and $j$; the probability of the event that they are not selected in one period is $1-\frac{2}{N(N-1)}<1$, and thus the probability of this event
occurring for infinitely many periods is 0.

Next, we show that information must be complete in any stable network. If ${\bf{g}}$ is connected, then clearly there is complete information. If ${\bf{g}}$ is unconnected and information is not complete, then there must exist two unconnected agents such that their beliefs on each other's type remain at the prior. When they are selected they would form a link, which implies that $({\bf{g}},B)$ is not stable, a contradiction.
\end{proof}

\begin{proof}[Proof of Lemma 2]
Consider the following selection path:
\begin{itemize}
\item{1.} Fix a high-value agent $i^*$. In the first $n_m+n_h-1$ periods, select $i^*$ and every other medium-value or high-value agent.
\item{2.} In the following $2n_l$ periods, select $i^*$ and every low-value agent twice consecutively.
\item{3.} In the following $\frac{n_h(n_h-1)}{2}$ periods, select every pair of high-value agents.
\item{4.} In the following $n_l(n_l-1)$ periods, select every pair of low-value agents twice consecutively.
\end{itemize}

If $n_m+n_h\geq 1$: under both complete and incomplete information, after step $1$, a star with $i^*$ as the center and all other low-value agents as the periphery would be formed. In step 2, under complete information no link would be formed; under incomplete information, between $i^*$ and every low-value agent, a link would first be formed and then severed in the next period. After step $3$, there would be a link between every pair of high-value agents. It is clear that under both complete and incomplete information, the network formed after step $3$ is stable.

If $n_m+n_h=0$: under both complete information, it is clear that no link ever forms. Under incomplete information, during step 4 a link would be formed and then severed between every pair of low-value agents. Therefore, under both complete and incomplete information, the empty network after step $4$ is stable.
\end{proof}

\begin{proof}[Proof of Proposition 2]
1: We already know from Theorem 1 that if $\mathbb{E}[f(x)]<c$, the network stays empty under incomplete information for any $\kappa$ and $\gamma(\infty)$. If $n_m+n_h\leq 1$, clearly the network stays empty under complete information for any $\kappa$ and $\gamma(\infty)$, and the first part of the statement is proved. If $n_m+n_h\geq 1$, from the proof of Lemma 2, we can construct a $\gamma(\infty)$ such that under complete information the formation process converges to some non-empty network. Therefore the maximal difference can be achieved.

2: The claim that the minimal difference can always be achieved is a direct result from Theorem 1.

If $n_h\geq 2$, in any stable network under complete and incomplete information, any pair of high-value agents must be linked. Thus the maximal difference cannot be achieved. If $n_l=0$, then the formation processes under complete and incomplete information would be the same, so again the maximal difference cannot be achieved.

If $n_h<2$ and $n_l>0$, first consider the following selection path when $n_m+n_h\geq \frac{c-\max_{\text{$i$ is low-value}}f(k_i)}{\delta c}+1$:
\begin{itemize}
\item{1.} Fix an agent $j^*\in\arg\max_{\text{$i$ is low-value}}f(k_i)$. In the first $n_m+n_h$ periods, select $j^*$ and every medium-value or high-value agent.
\item{2.} In the following $n_l-1$ periods, select $j^*$ and every other low-value agent twice consecutively.
\item{3.} In the remaining periods, let the selection path be the same as in the proof of Lemma 2.
\end{itemize}
Under complete information, as in the proof of Lemma 2, the formation process would converge to a network only consisting of links between medium-value or high-value agents. Under incomplete information, after step 1, a star with $j^*$ as the center and all the medium-value or high-value agents as the periphery would be formed. In step 2, a link would be formed and then severed between $j^*$ and every other low-value agent. After that, information becomes complete and no low-value agent except $j^*$ would ever be linked. For every medium-value or high-value agent, since $n_m+n_h\geq\frac{c-\max_{\text{$i$ is low-value}}f(k_i)}{\delta c}+1$, the benefit from the link with $j^*$ is at least $\max_{\text{$i$ is low-value}}f(k_i)+\delta(n_m+n_h-1)c\geq c$, which implies that the agent has incentive to maintain the link. In addition, as $n_h<2$, no link would be formed between any pair of medium-value or high-value agents, and thus the network is stable. This last fact also shows that there are no common links between the networks converged to under complete and incomplete information, and thus the maximal difference can be achieved.

Secondly, consider the following selection path when $n_m+n_h\geq 2$, $\delta$ is sufficiently close to 1 and $n_l\geq n_m+n_h-1$:
\begin{itemize}
\item{1.} In the first period, select a low-value agent and a medium-value or high-value agent; in the second period, select a second medium-value or high-value agent and the previous low-value agent; in the third period, select a second low-value agent and the previous medium-value or high-value agent; $\cdots$; in the $2(n_m+n_h-1)$th period, select the last medium-value or high-value agent and the previous low-value agent.
\item{2.} In the following $n_l-(n_m+n_h-1)$ periods, select a medium-value or high-value agent and every remaining low-value agent.
\item{3.} In the remaining periods, let the selection path be the same as in the proof of Lemma 2.
\end{itemize}
Under complete information, as in the proof of Lemma 2, the formation process would converge to a network only consisting of links between medium-value or high-value agents. Under incomplete information, after step 1, a line network only consisting of links between a low-value agent and a medium-value or high-value agent is formed. After step 2, information becomes complete, and as $\delta$ is sufficiently close to $1$, the network is stable (note that $\delta$ being sufficiently close to $1$ is consistent with the condition $n_h<2$). Therefore there are no common links between the networks converged to under complete and incomplete information, and thus the maximal difference can be achieved.
\end{proof}

\begin{proof}[Proof of Lemma 3]
Note that the social welfare is the sum of each agent's payoff. For agents having the same links in ${\bf{g}}_1$ and ${\bf{g}}_2$, it is clear that they are weakly better off in ${\bf{g}}_1$.

Now consider an agent $i$ whose links in ${\bf{g}}_1$ is a proper superset of those in ${\bf{g}}_2$. Let $ij_1,\cdots,ij_m$ denote $i$'s links in ${\bf{g}}_1$ but not in ${\bf{g}}_2$. Suppose that $u_i(k_i,k_{-i},{\bf{g}}_1)<u_i(k_i,k_{-i},{\bf{g}}_2)$. It implies that there must be a permutation of $ij_1,\cdots,ij_m$, denoted $ij'_1,\cdots,ij'_m$, such that for some $m'\in\{1,\cdots,m\}$, $u_i(k_i,k_{-i},{\bf{g}}_1-ij'_1-\cdots-ij'_{m'})>u_i(k_i,k_{-i},{\bf{g}}_1-ij'_1-\cdots-ij'_{m'-1})$.

Denote $L_i$ as an arbitrary proper subset of $i$'s links (including the empty set), and observe that for any ${\bf{g}}$ and any of $i$'s link $ij$, $u_i(k_i,k_{-i},{\bf{g}}-ij)-u_i(k_i,k_{-i},{\bf{g}})\geq u_i(k_i,k_{-i},{\bf{g}}\setminus L_i-ij)-u_i(k_i,k_{-i},{\bf{g}}\setminus L_i)$. Therefore, $u_i(k_i,k_{-i},{\bf{g}}_1-ij'_{m'})-u_i(k_i,k_{-i},{\bf{g}}_1)\geq u_i(k_i,k_{-i},{\bf{g}}_1-ij'_1-\cdots-ij'_{m'})-u_i(k_i,k_{-i},{\bf{g}}_1-ij'_1-\cdots-ij'_{m'-1})>0$, which implies that in ${\bf{g}}_1$, severing $ij'_{m'}$ would strictly increase $i$'s payoff. But this is a contradiction with the assumption of stability, and thus it must be the case that $u_i(k_i,k_{-i},{\bf{g}}_1)\geq u_i(k_i,k_{-i},{\bf{g}}_2)$. Therefore, we can conclude that ${\bf{g}}_1$ yields a weakly higher social welfare than ${\bf{g}}_2$.
\end{proof}

\begin{proof}[Proof of Proposition 3]
1: We already know from Theorem 1 that if $\mathbb{E}[f(x)]<c$, the network stays empty under incomplete information, yielding $W_{IC}(\kappa)=0$. Therefore, $W_C(\kappa)>W_{IC}(\kappa)$ if and only if there is some non-empty network in $G_C^*(\kappa)$, which is equivalent to the condition $n_m+n_h>1$.

2: The claim that $W_C(\kappa)\leq W_{IC}(\kappa)$ is a direct result from Theorem 2.

If $n_l=0$, $G_C^*(\kappa)$ and $G_{IC}^*(\kappa)$ are identical, and thus $W_C(\kappa)=W_{IC}(\kappa)$.

If $n_l>0$ and $n_m+n_h=1$, under complete information the network stays empty, yielding a social welfare of $0$. By Lemma 3, we know that in the stable wheel network, every agent's payoff is at least $0$. In addition, since $n_m+n_h=1$ the assumption that such a network is stable implies that the total number of agents is at least $5$, and that the medium-value or high-value agent must be non-singleton in this network. Thus, the two low-value agents who link with the medium-value or high-value agent must have a strictly positive payoff, which means that the social welfare is strictly positive. Finally, it is easy to see that this network can be formed under complete information. Thus $W_C(\kappa)<W_{IC}(\kappa)$.

If $n_l>0$ and $n_m+n_h>1$, consider the network ${\bf{g}}\in G_C^*(\kappa)$ which yields the highest social welfare (this network must exist, since there are only finitely many networks in $G_C^*(\kappa)$). Let $\delta$ be sufficiently close to $1$ such that ${\bf{g}}$ is minimal. Thus, there exist medium-value or high-value agents $i$ and $j$ such that $ij$ is the only link $i$ has in ${\bf{g}}$. Note that since ${\bf{g}}$ is minimal, ${\bf{g}}$ is also stable for any larger $\delta$.

Consider a selection path under complete information in which ${\bf{g}}$ emerges, such that no link is formed or severed after $ij$ is formed, and no low-value agent is selected before $ij$ is formed. Under incomplete information, consider the following variation of this selection path: before the period in which $ij$ is formed, insert two periods: in the first period, select some low-value agent $i'$ and $i$; in the second period, select $i'$ and $j$. Since $\mathbb{E}[f(x)]\geq c$, we know that $ii'$ and $i'j$ would both be formed. As $\delta$ gets sufficiently close to $1$, the payoffs of the medium-value or high-value agents would strictly increase due to the connection to $i'$. Therefore $W_C(\kappa)<W_{IC}(\kappa)$.
\end{proof}

\begin{proof}[Proof of Proposition 4]
We prove the result by construction. Consider the selection path that $i$ is selected twice consecutively with a low-value agent, then selected twice consecutively with another low-value agent, and so on. We know that initially a link forms and then breaks each time $i$ is selected with a different low-value agent. Let $p'_m$ be the posterior probability that $i$ is a medium-value or high-value agent after the link between $i$ and the $m$th low-value agent breaks. We know that by Bayesian updating, $p'_{m+1}=\frac{p'_m(1-p'_0)}{1-p'_mp'_0}$ with the initial condition that $p'_0$ is equal to the prior probability that an agent is medium-value or high-value. By assumption, we know that $p'_0<1$. Therefore $\frac{p'_{m+1}}{p'_m}=\frac{1-p'_0}{1-p'_mp'_0}\leq\frac{1-p'_0}{1-{p'_0}^2}<1$, and thus there exists a sufficiently large $n_l$ such that $\mathbb{E}[f(k_i)|\sigma(t)]<c$.
\end{proof}

\bibliographystyle{acm}
\bibliography{reference}
                             
\end{document}